\documentclass[aps,prx,twocolumn,superscriptaddress,noshowpacs]{revtex4-2}

\usepackage{times}

\usepackage{graphicx}
\usepackage{soul,color}
\usepackage{float}
\usepackage{bm}
\usepackage{braket}
\usepackage{placeins}
\usepackage{amsmath,amssymb}
\usepackage{comment}
\usepackage[dvipsnames]{xcolor}
\usepackage{url}
\usepackage[normalem]{ulem}
\usepackage{bm,nicefrac,xfrac}
\usepackage{physics}

\usepackage{xcolor}
\usepackage[normalem]{ulem}

\usepackage{xcolor}

\usepackage[normalem]{ulem}

\begin{document}

\title{Altermagnetic boosting of chiral phonons} 


\author{J. Okamoto}
\affiliation{National Synchrotron Radiation Research Center, Hsinchu 300092, Taiwan}

\author{C. Y. Mou}
\affiliation{Department of Physics, National Tsing Hua University, Hsinchu 300044, Taiwan}

\author{H. Y. Huang}
\author{G. Channagowdra}
\affiliation{National Synchrotron Radiation Research Center, Hsinchu 300092, Taiwan}

\author{C. Won}
\affiliation{Laboratory for Pohang Emergent Materials and Max Plank POSTECH Center for Complex Phase Materials, Department of Physics, Pohang University of Science and Technology, Pohang 37673, Korea}
\author{K. Du} 
\author{{X. Fang}} 
\affiliation{Keck Center for Quantum Magnetism and Department of Physics and Astronomy, Rutgers University, Piscataway, NJ 08854, USA}

\author{E.~V.~Komleva}
\affiliation{Institute of Metal Physics, 620041 Ekaterinburg GSP-170, Russia}
\affiliation{Department of Theoretical Physics and Applied Mathematics, Ural Federal University, 620002 Ekaterinburg, Russia}

\author{C. T. Chen}
\affiliation{National Synchrotron Radiation Research Center, Hsinchu 300092, Taiwan}

\author{S.~V.~Streltsov}
\affiliation{Institute of Metal Physics, 620041 Ekaterinburg GSP-170, Russia}
\affiliation{Department of Theoretical Physics and Applied Mathematics, Ural Federal University, 620002 Ekaterinburg, Russia}

\author{A. Fujimori}
\affiliation{National Synchrotron Radiation Research Center, Hsinchu 300092, Taiwan}
\affiliation{Department of Physics, National Tsing Hua University, Hsinchu 300044, Taiwan}
\affiliation{Department of Physics, University of Tokyo, Bunkyo-Ku, Tokyo 113-0033, Japan}

\author{S-W. Cheong}
\altaffiliation [email: ] {\emph{sangc@physics.rutgers.edu}} 
\affiliation{Keck Center for Quantum Magnetism and Department of Physics and Astronomy, Rutgers University, Piscataway, NJ 08854, USA}

\author{D. J. Huang}
\altaffiliation [email: ] {\emph{djhuang@nsrrc.org.tw}} 
\affiliation{National Synchrotron Radiation Research Center, Hsinchu 300092, Taiwan}
\affiliation{Department of Physics, National Tsing Hua University, Hsinchu 300044, Taiwan}
\affiliation{Department of Electrophysics, National Yang Ming Chiao Tung University, Hsinchu 300093, Taiwan}


\begin{abstract}
Chirality characterizes the asymmetry between a structure and its mirror image and underlies a wide range of chiral functionalities. In crystallographically chiral materials, phonons with non-zero linear momentum \textbf{k} can acquire a $k$-induced longitudinal magnetization, giving rise to chiral phonons. Helical spin order, with its proper screw-type configuration, breaks all mirror symmetries and therefore carries magnetic chirality. Such helical spins also generate non-relativistic spin splitting for any quasiparticle excitations propagating along the screw axis. To explore the possible connection between chiral phonons and magnetic chirality, we investigated the crystallographically polar and chiral compound (Mn,Ni)$_3$TeO$_6$, which hosts three distinct states: a paramagnetic state, a helical spin state with magnetic chirality, and a collinear spin state without magnetic chirality. We find an approximately tenfold enhancement of chiral-phonon coupling in the helical spin state along the screw axis, compared with both the paramagnetic and collinear spin states. These results identify a new route to amplify chiral phonons through an altermagnetic effect arising from the broken parity-time symmetry in helical spins. 
\end{abstract}

\date{\today}

\maketitle

\thispagestyle{empty}

Chirality captures the intrinsic asymmetry between a structure and its mirror image and plays a crucial role across diverse scientific disciplines \cite{YanAnnuRev2024}. 
It may preserve time-reversal symmetry while breaking all mirror symmetries. This concept also extends to elementary particles and quasiparticles as a quantum property. For example, in crystallographically chiral materials, phonons that propagate with finite linear momentum can develop a momentum-driven longitudinal magnetization, giving rise to chiral phonons~\cite{zhuScience2018,kishine2020,YinAdvMater2021,ishitoNatPhys2023,UedaNature2023,tsunetsuguJPSJ2023,OishiPRB2024,Wang2024,wangNanoLett2024}. Notably, chiral phonons consist of two enantiomeric modes that propagate with finite group velocity and are interconvertible by spatial inversion~\cite{ishitoNatPhys2023}.

Recently, chiral phonons have been extensively studied.
The concept of chiral phonons in chiral systems aligns perfectly with the hypothesis of the longitudinal kinetomagnetism of chirality, which postulates that any moving quasiparticle, such as a phonon or magnons in chiral systems, induces magnetization in its direction of motion, and consequently the quasiparticle itself acquires chirality due to the induced magnetization \cite{Furukawa2017,cheong2024APL,limPNAS2024}. In cyclotron motion of constituent atoms, the magnetic moments of chiral phonons are typically on the order of  $10^{-3}\mu_{\rm B}$ or less~\cite{Juraschek2019}, significantly smaller than those of electrons by a factor of the mass ratio between ions and electrons. This is purely owing to the orbital angular momentum of the atomic motion. However, recent experiments have shown that chiral phonons can be excited by intense laser pulses with circular polarization ~\cite{Luo2023} and have large effective magnetic moments~\cite{Cheng2020,Lujan2024}. 

Despite growing interest, research on chiral phonons remains rather scattered, with no essential guidelines for systematically enhancing their functionalities. Yet, to the best of our knowledge, no experimental evidence has demonstrated a strong coupling between chiral phonons and spin structures. As a result, their potential for functional applications has remained largely unexplored. A crucial open question concerns the interplay between chiral phonons and magnetic chirality.
A helical-spin lattice itself breaks all mirror symmetries \cite{cheongnpjQM2022}. This magnetically chiral state breaks parity-time~($\mathcal{PT}$) symmetry while preserving time-reversal ($\mathcal{T}$) symmetry, corresponding to the A-type altermagnet in the refined definition of altermagnetism~\cite{cheong2025}, in which $\mathcal{T}$ symmetry remains intact whereas $\mathcal{P}$ symmetry is broken. Such a system has non-relativistic spin splitting along the screw axis \cite{calvo1979,zadorozhnyi2023}, and kinetomagnetism is intrinsically present.  
Therefore, understanding how chiral phonons interact with magnetic chirality
is of great importance, as changes in phonon chirality induced by magnetic chirality may occur through an altermagnetic effect.
Crystallographically chiral and polar Ni$_3$TeO$_6$ \cite{Zivkovic2010,OhNatComm2014,WangAPLM2015,Okamoto2024} exhibits three distinct magnetic states when doped with Mn: a paramagnetic phase, a helical spin state with magnetic chirality, and a collinear spin state without magnetic chirality \cite{KimPRM2021}. This diversity makes (Mn,Ni)$_3$TeO$_6$ an ideal platform for investigating the coupling between structural chirality and electron spin. 

\begin{figure}[t]
\centering 
\includegraphics[width=1\columnwidth]{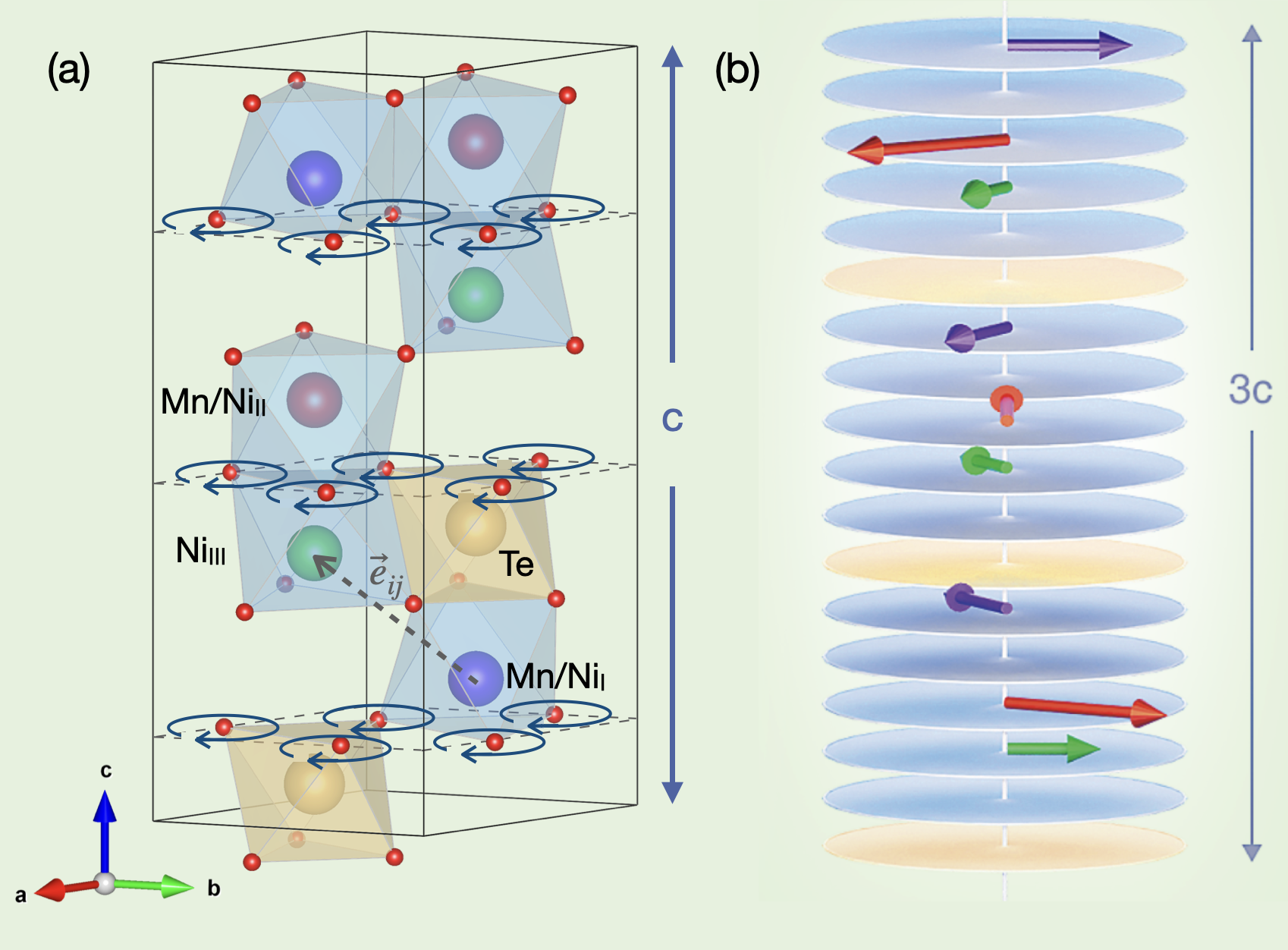}
\caption{{\bf Cartoon illustration of chiral phonons induced by helical spins.} 
(a) crystal  structure of Mn-doped Ni$_3$TeO$_6$ and illustration of its chiral phonons (the collective rotation of O ions is predominantly but not totally in the $ab$-plane and is shown for the second mode from the top of the spectrum given in Fig.~\ref{DFT}). Ni$_3$TeO$_6$ crystallizes in a noncentrosymmetric corundum structure with three nonequivalent Ni sites. The crystal contains two kinds of honeycomb layers formed by edge-sharing NiO$_6$ and TeO$_6$ octahedra. Ni$_{\rm I}$O$_6$ and Ni$_{\rm II}$O$_6$ are ferromagnetically coupled and form one honeycomb layer, while Ni$_{\rm III}$O$_6$ and TeO$_6$ form the other. These honeycomb layers stack along the $c$-axis, generating a polar crystal structure \cite{Zivkovic2010,OhNatComm2014,WangAPLM2015}. Additionally, the relative positions of these octahedra in the $ab$-plane revolve along the $c$-axis, demonstrating the handedness of crystal chirality. When Mn is doped into Ni$_3$TeO$_6$, the Mn ion predominantly occupies the Ni$_{\rm I}$ and Ni$_{\rm II}$ sites \cite{KimPRM2021}. (b) Illustration of helical spins. Mn-doped Ni$_3$TeO$_6$ shows an incommensurate helical spin order between 60~K and 74~K. The arrows, with lengths indicating their relative magnitudes, represent the Ni or Mn spins, which are ferromagnetically aligned in each honeycomb layer, as depicted by colored disks. The modulation vector of helical spins was determined as $\vb{k} = (0, 0, 1.5{\pm}\delta)$ with $\delta = 0.146$ in r.l.u. \cite{KimPRM2021}.}
\label{structure}
\end{figure}

In this Article, we present high-resolution O $K$-edge resonant inelastic X-ray scattering (RIXS) measurements on Mn-doped Ni$_3$TeO$_6$ to address the non-relativistic enhancement of chiral phonons. 
The RIXS spectra reveal chiral-phonon excitations at 85 meV, displaying pronounced circular dichroism exclusively in the helical-spin phase. Before the magnetic order sets in, chiral phonons already exist due to the crystal chirality \cite{WangAPLM2015,Okamoto2024}. Once the helical spin order emerges, the system exhibits strong altermagnetism, as spin-split bands arise from exchange coupling in the non-relativistic limit (i.e., in the absence of SOC). Consequently, 
chiral phonons are enhanced through the kinetomagnetic coupling between phonon chirality and the helical spin structure, and this enhancement is inherently non-relativistic. Upon further cooling into the collinear spin phase, where spin splitting appears only in the presence of finite SOC, the enhancement is significantly reduced.

\begin{figure}[t]
\includegraphics[width=1\columnwidth]{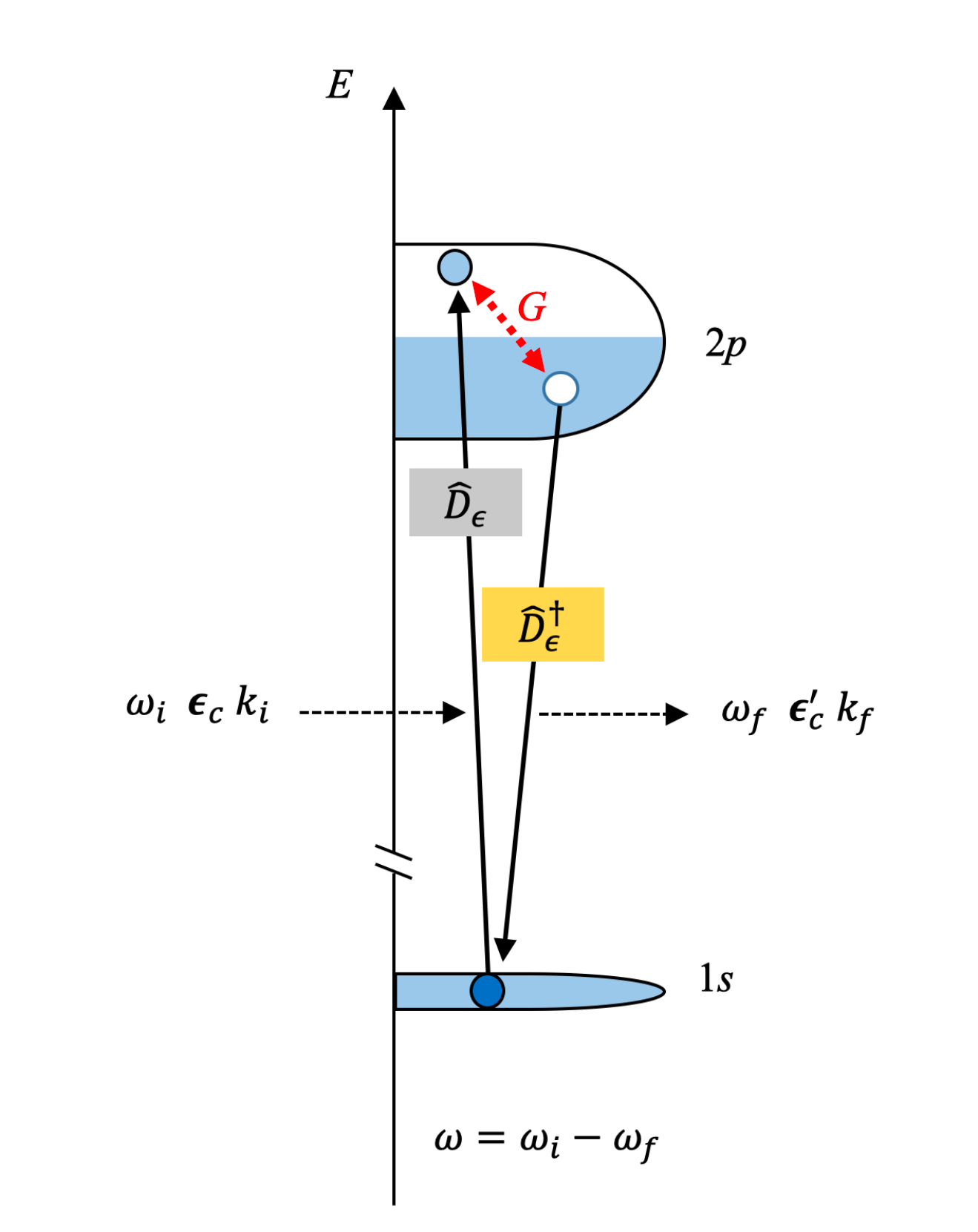}
\caption{{\bf Schematic diagram of O $K$-edge RIXS with circular polarization.} This diagram illustrates the transitions involved in the O $K$-edge RIXS. Circularly polarized X-rays with energy $\omega_i$, polarization $\bm{\epsilon}_c$, and wavevector $\bm{k}_i$ excite an electron from the $1s$ core shell into an unoccupied state in the O $2p$ band through the dipole operator $\hat{D}_{\epsilon} ={\bm{\epsilon}_c}{\cdot}\bm{p}^{\dagger}s$, where $\bm{p}^{\dagger}$ and $s$ are creation and annihilation operators of the $2p$ and $1s$ electrons, respectively. The intermediate state is governed by a propagator $G$ determined by the ground-state Hamiltonian $H_0$ and the intermediate-state Hamiltonian $H_{\rm I}$, and the inverse core-hole lifetime $\Gamma$. The system then relaxes to the RIXS final state through $\hat{D}_{\epsilon}^{\dagger}$, emitting X-rays with energy $\omega_f$, polarization $\bm{\epsilon}'_c$, and wavevector $\bm{k}_f$. The energy loss is given by $\omega = \omega_{i} - \omega_f$. The electron-phonon coupling in RIXS leads to a measurement of phonon excitations \cite{AmentEPL2011,geondzhianPRB2020,DashwoodPRX2021,okamotonpjQM2025}. For chiral phonons, the RIXS circular dichroism, i.e., the contrast between RCP and LCP, results from the circular polarization of phonons.} 
\end{figure}

\begin{figure}[h]
\centering 
\includegraphics[width=1\columnwidth]{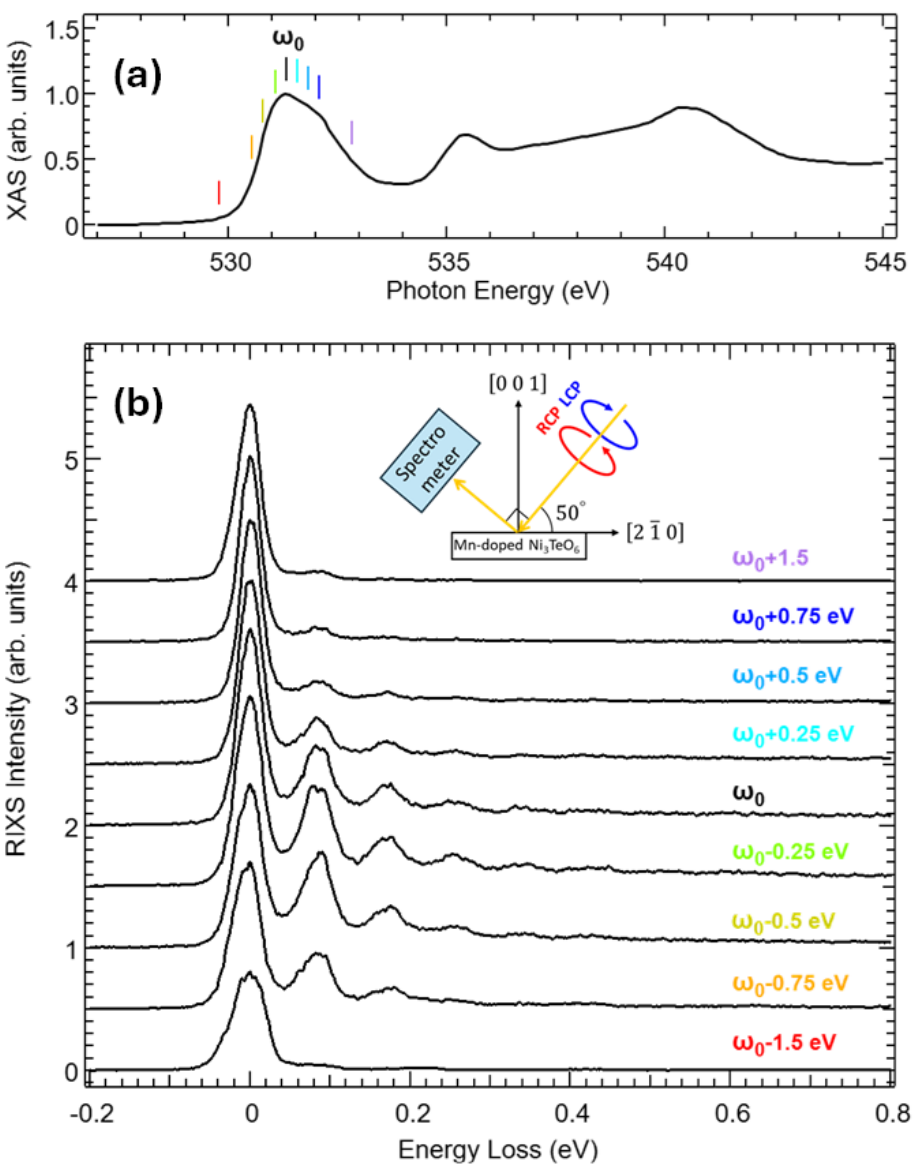}
\caption{{\bf O $K$-edge XAS and photon-energy dependent RIXS of Mn-doped Ni$_3$TeO$_6$.}
(a) X-ray absorption spectrum measured using fluorescence-yield, RCP, and normal incidence relative to the $ab$ plane at 50 K. $\omega_{0}$ is XAS peak at 531.3 eV. (b) RIXS spectra measured using LCP at 50 K for the incident photon energies indicated by the vertical colored bars shown in (a). Spectra are vertically shifted for clarity. The inset illustrates the experimental geometry. The sample surface was the (001) plane. The scattering plane was in the $ac$ plane of the crystal, while the incident and the scattering angles were 50$^{\circ}$ and 90$^{\circ}$, respectively.} 
\label{OKXASRIXS}
\end{figure}

\begin{figure*}[t]
\centering 
\includegraphics[width=2\columnwidth]{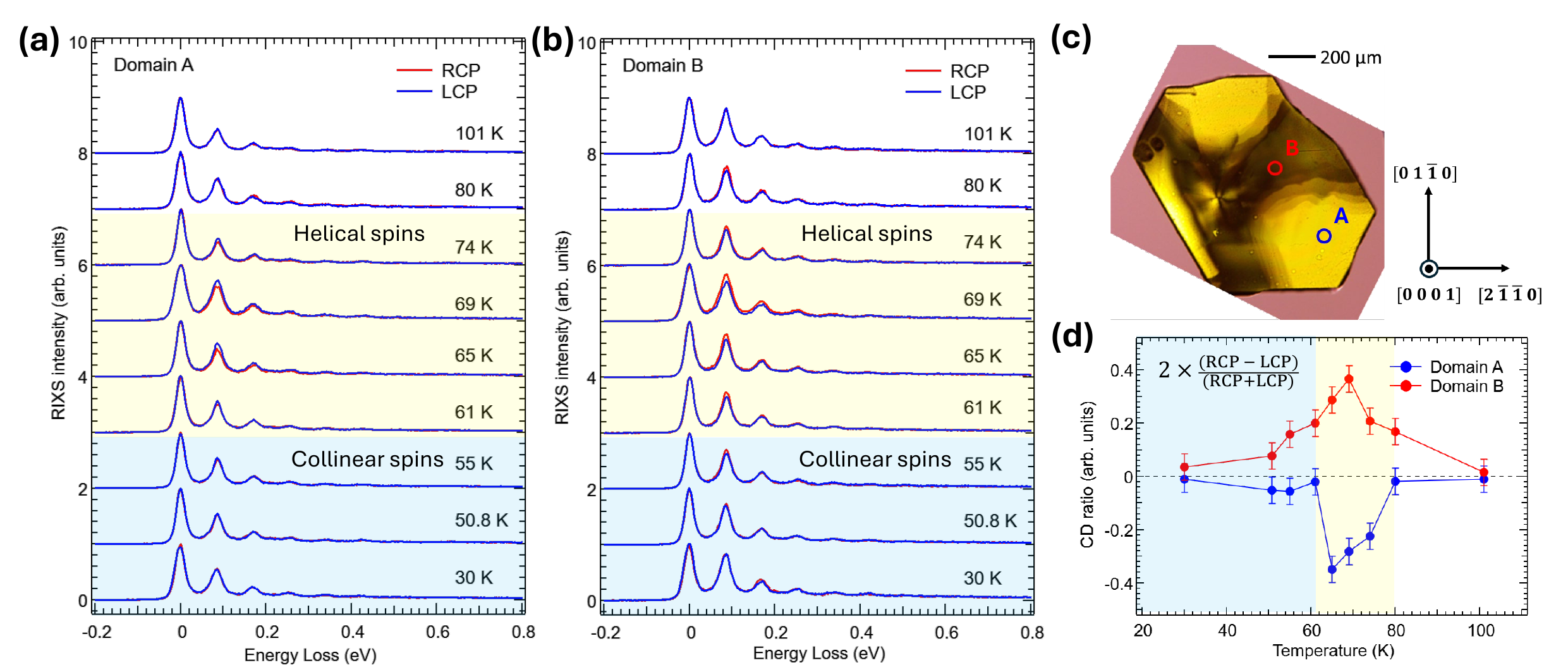}
\caption{{\bf Circular dichroism in chiral phonons of Mn-doped Ni$_3$TeO$_6$.}
(a) \& (b) O $K$-edge RIXS of Mn-NTO sample measured with incident photon energy of $\omega_{0}-0.5$ eV and $q = (-0.027, 0.0135, 0.83)$ with RCP (red) and LCP (blue) at various temperatures at domain A(a) and B(b). RIXS spectra are normalized to the elastic peak intensity. Yellow and blue background indicate the antiferromagnetic (AFM) phases with helical spins (yellow) and collinear spins (blue). Clear difference between RCP and LCP, i.e. circular dichroism (CD) in phonon structures are observed at helical AFM phase though not or negligibily small at paramagnetic and collinear AFM phase. Sign of CDs are reversed between opposite chirality domains; negative (positive) in the domain A(B).
(c) Transmission polarized optical microscope image of the sample. Bright and dark regions correspond to the domains of opposite chiralities: A (blue) and B (red) marks are the measured position of each domain and diameters of the marks show the incident beam size of 50 $\mu$m. (d) Temperature dependence of the CD ratio $2\times\frac{\rm RCP-LCP}{\rm RCP+LCP}$ of the first phonon structure at $\sim$90 meV. Strong CD as $\sim$30 \% were observed in the helical AFM phase though those in the paramagnetic and collinear AFM phases were negligibly small as less than 5 \%.} 
\label{OKXASRIXS2}
\end{figure*}

\vspace{5mm}
\noindent{\Large\bf Results}
\vspace{2mm}

\noindent {\bf Mn-doped Ni$_3$TeO$_6$ with helical spins}. Nickel tellurite, Ni$_3$TeO$_6$, crystallizes in a modified corundum structure consisting of three Ni sites and one Te site per formula unit; see Fig.~\ref{structure}(a). In this structure, two types of honeycomb layers—one formed by edge-sharing Ni$_{\rm I}$O$_6$ and Ni$_{\rm II}$O$_6$ octahedra, and the other formed by edge-sharing Ni$_{\rm III}$O$_6$ and TeO$_6$ octahedra—are alternately stacked along the $c$-axis, resulting in a polar and chiral crystal structure. The Ni$_{\rm I}$ and Ni$_{\rm II}$ spins are ferromagnetically ordered, while Ni$_{\rm III}$ is ferromagnetically (antiferromagnetically) coupled to Ni$_{\rm I}$ and Ni$_{\rm II}$ above (below) the Ni$_{\rm I}$-Ni$_{\rm II}$ layer. Additionally, the relative positions of these octahedra in the $ab$ plane revolve along the $c$-axis, giving rise to the chiral handedness of the crystal  \cite{WangAPLM2015, Okamoto2024}.

When Mn is doped into Ni$_3$TeO$_6$, the Mn ion predominantly occupies the Ni$_{\rm I}$ or Ni$_{\rm II}$ site \cite{KimPRM2021}. Upon cooling to the Neel temperature $T_{N_1} = 74$~K, the long-range antiferromagnetic order emerges, with magnetic moments aligned in the $ab$ plane for temperatures between $T_{N_1}$ and $T_{N_2} = 60$~K \cite{KimPRM2021}. In this intermediate phase, an incommensurate helical spin order emerges, characterized by a wavevector $\vb{k} = (0, 0, 1.5{\pm}\delta)$ propagating along the $c$-axis with $\delta = 0.146$ in reciprocal lattice units (r.l.u.), as illustrated in Fig.~\ref{structure}(b). Below $T_{N_2}$, the spins align collinearly along the $c$-axis with the Ising-type anisotropy, similar to the behavior observed in pristine Ni$_3$TeO$_6$ \cite{Zivkovic2010,OhNatComm2014}. This unique system facilitates competition between collinear and helical magnetic structures with minimal chemical disorder, providing an ideal platform to investigate the kinetomagnetism of magnetic chirality.

\vspace{5mm}
\noindent {\bf Circular dichroism of phonons}. We used O $K$-edge RIXS to measure the circular dichroism in the phonon excitations of Mn$_x$Ni$_{3-x}$TeO$_6$ with $x$ = 0.6-0.7 at various temperatures.  
As illustrated in Fig. 2, O $K$-edge RIXS involves the excitation of a core-level $1s$ electron to unoccupied $2p$ states through the dipole operator $\hat{D}_\epsilon$, when the incident X-ray energy matches the energy difference between these states \cite{AmentRMP2011,deGroot2024RIXS}. After interactions in the intermediate state governed by a propagator $G$, the system relaxes to its final state. RIXS detects phonons through electron-phonon coupling in the intermediate state \cite{AmentEPL2011,geondzhianPRB2020,DashwoodPRX2021,okamotonpjQM2025}, and the use of circularly polarized X-rays enables magnetization-sensitive measurements. RIXS has been proven to be effective in detecting chiral phonons using circularly polarized X-rays \cite{UedaNature2023}. Before performing RIXS, we first measured the X-ray absorption spectrum (XAS). Figure 3(a) shows the O $K$-edge XAS of Mn-doped Ni$_3$TeO$_6$ at 50~K; the lowest-energy XAS peak was at energy $\omega_{0} = 531.3$~eV, corresponding to the transition of O $1s$ to the $2p$ states above the Fermi level. Figure 3(b) displays the RIXS spectra with momentum transfer around $q = (-0.027, 0.0135, 0.83)$ excited by various X-ray energies across $\omega_0$, showing phonon excitations centered at 85~meV with harmonics up to the fifth order. The observed phonon energy agrees with the results of Raman scattering \cite{BlasseJSSC1972,Skiadopoulou2017,RetuertoPRB2018}. 



The RIXS intensity of the phonon excitations was maximum when the incident X-ray was 0.5~eV below $\omega_{0}$. We therefore set the incident X-ray energy of RIXS to $\omega_{0}-0.5$~eV for the circular dichroism measurements. Although chiral phonons exist in non-magnetic phases because of crystal chirality, their circular dichroism remains beyond the sensitivity of our RIXS measurements. In the paramagnetic phase above $T_{\rm N}$, the RIXS spectra obtained with right-handed (RCP) and left-handed (LCP) circularly polarized X-rays overlapped well. In contrast, we found that, as shown in Fig. 4(a), the phonon excitations in RIXS with RCP and LCP displayed a noticeable contrast, i.e., circular dichroism, for a given crystal chirality (handedness) above 65~K and below $T_{\rm N}$. The sign of this circular dichroism reversed when the crystal chirality was reversed, as shown in Fig. 4(b). In other words, the handedness of the helical spins controls the sign of the phonon circular dichroism. Interestingly, when the temperature was cooled below 60~K, this circular dichroism vanished in the collinear antiferromagnetic phase. The fact that the circular dichroism of phonon excitations is solely enhanced in the helical-spin phase indicates that chiral spins significantly amplify the chiral phonons that already exist in a chiral crystal.

\begin{figure}
    \centering
    \includegraphics[width=1.0\linewidth]{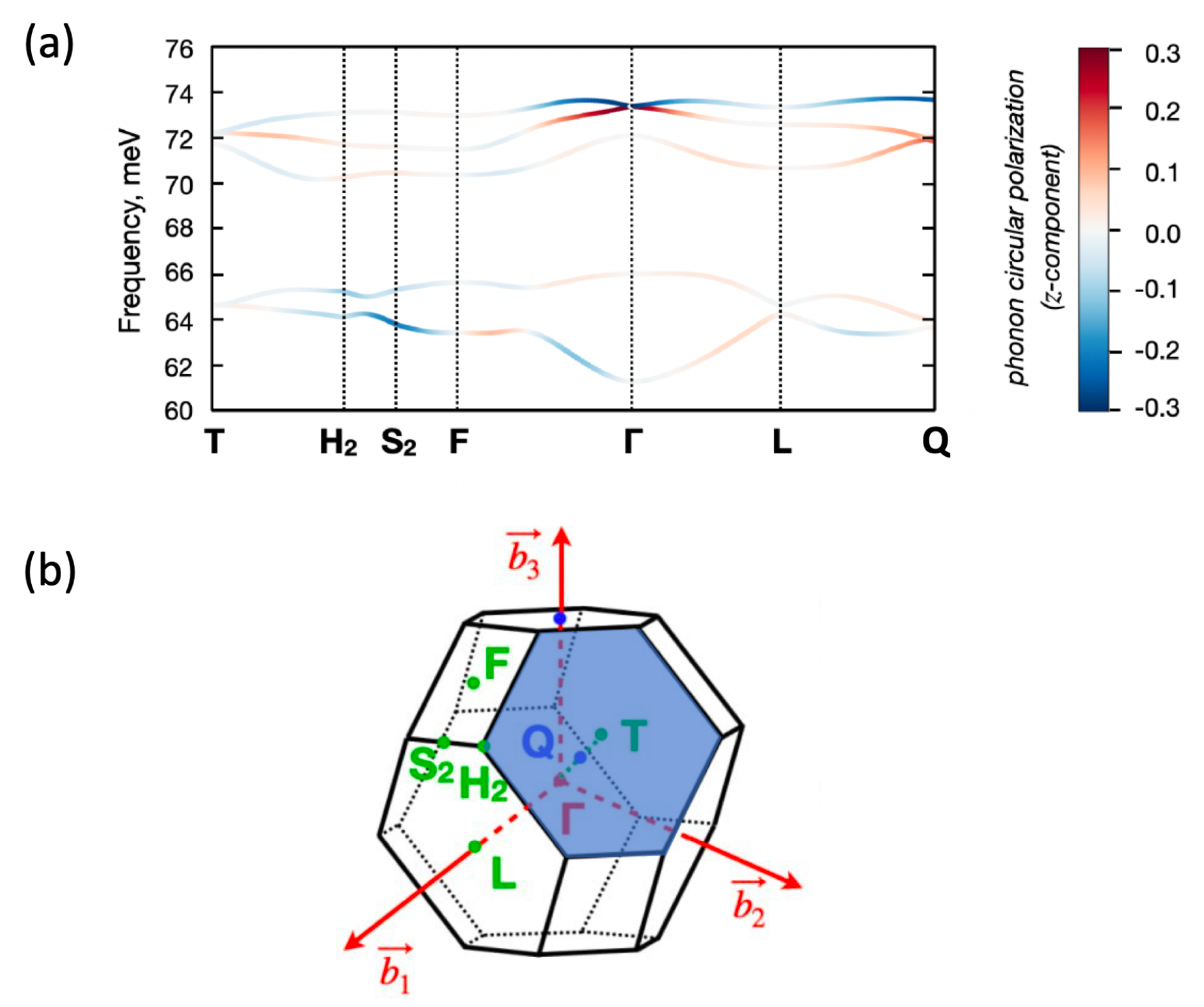}
\caption{{\bf DFT results of the lattice dynamics and phonon chirality simulations.}
(a) The phonon dispersion for the high-energy modes. The spectra are given for the primitive cell and the collinear antiferromagnetic. 
The color scale bar indicates the calculated value of the $z$-component of the circular phonon polarization for each mode. (b) The Brillouin Zone of the primitive cell of Ni$_3$TeO$_6$. The momentum transfer ${\bf Q}$ in RIXS measurements is (0.263 0.290 0.277) in units of the reciprocal lattice for the primitive cell.} 
    \label{DFT}
\end{figure}

\vspace{3mm}
\noindent {\bf DFT results.} We used {\it ab initio} calculations to study chirality in the phonon subsystem of Mn-doped Ni$_3$TeO$_6$. Strong electronic correlations were taken into account via DFT+U approach without spin-orbit coupling (SOC) (for details see Methods section), Mn-doping was introduced by changing the number of electrons. The collinear ferrimagnetic structure with spins parallel to the $c$ axis was considered (Ni$_{\mathrm{I}}$ \textuparrow, Ni$_{\mathrm {II}}$ \textdownarrow, Ni$_{\mathrm {III}}$ \textdownarrow). Energies of the high-frequency ligand-related vibrations are known to be often underestimated. In our case, theoretical phonon energies $\varepsilon_{ph} (\bm{k})$ are also $\sim$15\% lower than experimental ones as one can see from Fig.~\ref{DFT}(a). However, our calculations confirm that these are indeed high-frequency oxygen modes, which bear a substantial (phonon) orbital momentum (shown by color contrast) and are chiral.

There are two phonon modes with slightly different frequencies having strong contrast at experimental $\bm{Q}$ vector. 
Figure~\ref{structure}(a) illustrates corresponding vibrations of one of these modes, they involve chiral rotations (predominantly in the $ab$ plane) of oxygen ions. For the second phonon mode, ligands rotate in the opposite direction; see Fig. S1 in the Supplementary Information for animations of both phonon modes. 
Importantly, these vibrations can affect the exchange coupling both along the $c$ axis and in diagonal direction. This makes the situation in Ni$_3$TeO$_6$ more complex and rich than that in non-magnetic quartz \cite{UedaNature2023}. 
In next section, we develop a microscopic theory taking into account the presence of SOC to explain the observed enhancement of circular dichroism in phonon excitations. 


\vspace{3mm}
\noindent {\bf Microscopic model}. To understand the circular dichroism observed in RIXS, we begin with the Kramers-Heisenberg formula of the RIXS intensity $I(\omega)$ with $\omega$ being the excitation energy. For a RIXS process in which a ground state $\ket{0}$ with energy $E_0$ is excited to an intermediate $\ket{m}$ with energy $E_m$, and then relaxes to a final state $\ket{f}$ with energy $E_f$, the RIXS intensity is $I(\omega)=\sum_{f}\left| A_{f} \right|^{2}\delta({\omega}-E_{f}+E_{0})$, where $A_{f}$ is the scattering amplitude. 

Ni$_3$TeO$_6$ has a chiral crystal structure. 
For the phonon associated with the twisting of (Mn$_x$Ni$_{1-x}$)O$_6$ octahedra, we consider the rotation of the oxygen atom in the $ab$ plane about the $c$-axis ($z$-axis) with a rotation angle $\phi$. With the quantum pendulum as the ground state Hamiltonian \cite{UedaNature2023}, $H_{0} = L^{2}_{z}/2I + v\cos{\phi}$, where $I$ is the moment of inertia, $L_{z} = -i{\partial}_{\phi}$ is the angular momentum operator, and a negative potential $v <0$ confines the oscillation to around $\phi = 0$.

In O $K$-edge RIXS, $1s$ electrons are excited into the $2p$ orbitals. The excited $2p$ electrons will interact with the rest of the system through the intermediate Hamiltonian $H_{\rm I} = H^{\alpha}_{\rm I} +H^{\rm DM}_{\rm I} $, where $H^{\alpha}_{\rm I}$ accounts for charge scattering and $H^{\rm DM}_{\rm I}$ captures the Dzyaloshinskii–Moriya (DM) interaction due to  $3d$ spins of Ni or Mn \cite{Dzyaloshinsky1958,Moriya1960}. Using the Kramers-Heisenberg formula to calculate the RIXS scattering amplitude within the ultrashort core-hole lifetime approximation, we can prove that the RIXS scattering amplitude involving a transition from the ground state $\ket{0}$  to $\ket{f}$ is given by
\begin{equation}
\begin{aligned}
A_{f}\propto &\left[\alpha+\left(\bm{S}_{q}{\cross}\bm{S}_{-q}\right)_{z}(\lambda_{2}-\lambda_{1})\delta u_{0}\right]{\cross}\\
&\bra{f}\bm{\epsilon}'^{*}_{c}
\begin{pmatrix}
     0 & e^{-i2\phi}   \\
     e^{i2\phi}  & 0 
\end{pmatrix}
\bm{\epsilon}_{c}\ket{0},
\end{aligned}
\end{equation}
where $\bm{S}_{\bm{q}}$ is the magnetization at the modulation vector $\bm{q}$. The polarization $\bm{\epsilon}$ is expressed in terms of the circular polarization basis $\bm{\epsilon}_c = (\epsilon_{x}+i\epsilon_{y},~\epsilon_{x}-i\epsilon_{y})$, with $\bm{\epsilon}_c = (1,~0)$ and $(0,~1)$ corresponding to LCP and RCP, respectively. 
$\lambda_1$ and $\lambda_2$ are coupling constants for electrons in orbitals pointing toward the center of rotation and along the tangential direction, respectively. $\alpha$ scales with the electron-phonon coupling energy, and $\delta{u}_{0}$ is the displacement caused by the chiral spins through the DM interaction. Details of the derivation are provided in the Supplementary Information; see Eq.~(S26). 

This expression of the scattering amplitude explains the observed CD in RIXS. If the circular polarization of the X-ray photon in the scattering process changes from LCP to RCP, or vice versa, an angular momentum of ${\pm}2\hbar$ is transferred to phononic excitations, resulting in chiral phonons.  We note in passing that the broken parity symmetry in the helical spin state induces an additional spin–spin interaction as a \emph{non-relativistic} DM interaction. This interaction arises from the super-exchange between two neighboring spins whose quantization axes differ due to the helical structure: when super-exchange couples spins defined in different local frames, it naturally generates a DM-like term. This non-relativistic DM interaction complements the conventional, relativistic DM interaction that originates from spin–orbit coupling. In the helical spin state, the helical structure itself can be viewed as a condensate of chiral phonons. Consequently, the induced non-relativistic DM interaction not only enhances the dynamical coupling between chiral phonons and magnetic chirality but also reinforces the condensate by further increasing the static displacement $\langle\delta u_0\rangle$.


\vspace{5mm}
\noindent {\Large\bf Discussion}
\vspace{2mm}

We observed giant CD in RIXS excitations arising from bond-bending phonons associated with the twisting of (Mn$_x$Ni$_{1-x}$)O$_6$ octahedra in the helical-spin phase. These phonons are already chiral in the nonmagnetic phase due to the chiral crystalline structure. Although it lies below our detection limit due to insufficient energy resolution, the phonon spectrum in RIXS excited by circular X-rays in a chiral crystal is expected to exhibit circular dichroism originating from the crystal's chirality, similar to that observed in quartz \cite{UedaNature2023}. In contrast, these chiral phonons become strongly polarized only in the helical-spin phase. Microscopically the DM interaction and the inverse DM effect provide an efficient coupling between lattice and spin subsystems in a non-collinear phase, very similar to what happens in (spiral type-II) multiferroics~\cite{Katsura2005,Mostovoy2006,Khomskii2009}. 

The experimental data, together with the theoretical modeling, demonstrate that the non-collinear spin configuration produces a pronounced circular dichroism in the phonon spectrum through the intermediate helical state. Since  O $K$-edge RIXS is not sensitive to spin-flip excitations, the spin configuration in the low-energy RIXS excitations is effectively identical to that in the ground state. Hence, one can conclude that the system’s non-collinear spins amplify the CD of the chiral phonons. In other words, magnetic chirality enhances the chiral-phonon CD in RIXS. This explains why the CD in our RIXS experiments is related to spin chirality, while collinear spins do not enhance CD in the RIXS excitation of phonons. 

The observation of pronounced CD in phonon excitations indicates that chiral phonons are boosted. Here, chiral-phonon “boosting” carries two implications. First, a phonon branch that is intrinsically chiral should remain chiral, while branches that were originally achiral can acquire chirality through kinetomagnetism of chirality. Second, boosting also refers to enhancing the chiral-phonon condensate itself. Since chiral phonons are bosons, the emergent chiral structure we observe can be viewed as the macroscopic manifestation of a bosonic condensate. In this sense, boosting corresponds to increasing  $\langle\delta u_0\rangle$ of the underlying phonon condensate, and this boosting originates from SOC.

From the perspective of kinetomagnetism of chirality, the motion of chiral phonons in a helical magnetic structure significantly enhances their polarization. This enhancement aligns with the hybridization between spin excitations (chiral magnons) and chiral phonon excitations~\cite{cui2023}. In Ni$_3$TeO$_6$, significant magnon-phonon coupling was already reported via measurements of thermal Hall conductivity \cite{YangPRB2022}. In Mn-doped Ni$_3$TeO$_6$, chiral magnons generated by the magnetically chiral structure further enhance the CD in phonon excitations. This exchange-driven pathway provides a significantly strong coupling between phonon chirality and spin textures. When chiral phonons coexist with helical magnetic order, the phonon polarization is effectively boosted by a non-relativistic channel since the spin splitting in the helical spin order is generated by exchange interaction in the non-relativistic limit~\cite{calvo1979,zadorozhnyi2023}. In contrast, the collinear spin state below $T_{N_2} = 60$~K exhibits spin-split bands only in the presence of non-zero SOC, resulting in weak altermagnetism; therefore, the circular dichroism in phonon excitations is not enhanced. The observed enhancement is an \emph{altermagnetic effect} because the helical spin system is a strong altermagnet, breaking $\mathcal{PT}$ symmetry while exhibiting zero net magnetization. It represents a natural extension of non-relativistic spin splitting inherent to altermagnetism and opens new possibilities for controlling chiral excitations.





\vspace{5mm}
\noindent {\Large\bf Methods}
\vspace{2mm}

\noindent{\textbf{Sample synthesis}. Single crystals of Mn-doped Ni$_3$TeO$_6$ were grown by chemical vapor transport technique with polycrystalline powder and TeCl$_4$ as transport agent. In detail, polycrystalline Ni$_{3-x}$Mn$_{x}$TeO$_6$ was prepared using conventional solid state reactions at 650-750~$^{\circ}$C in air with several intermediate grindings. To obtain a high Mn concentration close to the solubility limit of Mn, the nominal Mn concentration was above 1. The evacuated quartz ampule containing polycrystalline Mn-doped Ni$_3$TeO$_6$ and TeCl$_4$ was maintained between 800~$^{\circ}$C and 730~$^{\circ}$C for 2 weeks. The Mn concentration in the grown single crystal was determined to be $x=0.65{\pm}0.05$ using X-ray diffraction and energy-dispersive X-ray spectroscopy. 

\vspace{5 mm}
\noindent\textbf{RIXS measurements}. O $K$-edge RIXS measurements were conducted at the AGM-AGS spectrometer of Beamline 41A of the Taiwan Photon Source \cite{SinghJSR2021}. 
The design of this AGM-AGS spectrometer is based on the energy compensation principle of grating dispersion. The energy bandwidth of the incident X-ray was 0.1 eV, while keeping the total energy resolution of RIXS at 25 meV. The XAS spectrum was measured in the fluorescence yield (FY) mode using a photodiode with normal-incident X-ray to the $ab$ plane of the crystal. The scattering plane of RIXS was in the $ac$ plane of the crystal. For RIXS measurements, the incidence angle was $50^{\circ}$ from the sample surface, and the scattering angle was fixed at 90$^{\circ}$.

\vspace{5 mm}
\noindent\textbf{DFT calculations}. {\it Ab initio} DFT calculations were performed in the VASP code \cite{PhysRevB.54.11169} while for the lattice dynamics simulations {\it via} frozen phonon method the program Phonopy \cite{phonopy-phono3py-JPSJ} was used. On-site electronic correlations were taken into account {\it via} DFT+U method in a form introduced by Dudarev {\it et al.} \cite{PhysRevB.57.1505} with the effective $U=$6.5~eV. 6$\times$6$\times$6 k-mesh for the Brillouin zone integration was chosen and the cut-off energy for the plane-wave basis was 700 eV.  Self-consistent calculations were carried out with the Perdew-Burke-Ernzerhof (PBE) version of the generalized gradient approximation (GGA) exchange-correlation functional \cite{PhysRevLett.78.1396}.
The crystal structure was taken from \cite{Ivanov2013} and the primitive cell with one formula unit was used. The 2$\times$2$\times$2 supercell with the corresponding reduction of the k-mesh density down to 4$\times$4$\times$4 was taken for the phonon calculations.
For chirality calculations the algorithm introduced in \cite{PhysRevLett.115.115502} was used. Animations of the chiral modes are presented in a form given by the TSS Physics project \cite{ChMode}.   

\vspace{5mm}
\noindent\textbf{Acknowledgements}. 
SVS would like to thank S. Artyukhin and P. van Loosdrecht for useful discussions. This work was supported in part by the Taiwan National Science and Technology Council under Grant Nos. NSTC112-2112-M-007-031 and NSTC113-2112-M-007-033. This work was also supported by the W. M. Keck foundation grant to the Keck Center for Quantum Magnetism at Rutgers University. We are thankful for the support of the Japan Society for the Promotion of Science under Grant No. JP22K03535. CW was supported by the National Research Foundation of Korea(NRF) funded by the Ministry of Science and ICT (No. RS-2022-NR068223). AF acknowledges the support of the Yushan Fellow Program under the Ministry of Education (MOE) of Taiwan. CYM acknowledges support from the Center for Quantum Science and Technology within the framework of the Higher Education Sprout Project by the MOE of Taiwan. EVK and SVS thank the Ministry of Science and Higher Education of the Russian Federation for supporting theoretical calculations through funding the Institute of Metal Physics. The work of EVK was supported by the grant from the Foundation for the Development of Theoretical Physics and Mathematics ``BASIS''.

\vspace{5mm}
\noindent\textbf{Author contributions}. DJH coordinated the project. JO, HYH, GC, DJH, and CTC conducted the RIXS experiments. CW, KD, XF, and SWC synthesized and characterized the samples. JO and DJH analyzed the RIXS data and prepared the figures. CYM developed the theoretical model of RIXS. EVK and SVS performed DFT calculations. DJH, CYM, SWC, SVS, and AF wrote the manuscript with inputs of other authors.

\vspace{5mm}
\noindent\textbf{Competing interests}. The authors declare that they have no competing interests.

\vspace{5mm}
\noindent\textbf{Data availability}. All raw data are plotted in the main text or the Supplementary Information. Their numeric values are available from the corresponding author upon reasonable request.

\bibliography{ref_MnNTO}



\end{document}


\title{{\large Supplementary Information for} \\ Altermagnetic boosting of chiral phonons}

\author{J. Okamoto, C. Y. Mou, H. Y. Huang, G. Channagowdra, Choongjae Won, \\ Kai Du, Xiaochecn Fang, E. V. Komleva, C. T. Chen, S. V. Streltsov, \\ A. Fujimori, S-W. Cheong, and D. J. Huang}

\date{\normalsize\today}

\maketitle

\hypersetup{%
    pdfborder = {0 0 0}
}

\renewcommand{\listfigurename}{List of plots}
\tableofcontents


\newpage

\section{O K-edge RIXS}


For a RIXS process in which a ground state $\ket{0}$ with energy $E_0$ is excited to an intermediate $\ket{m}$ with energy $E_m$, and then relaxes to a final state $\ket{f}$ with energy $E_f$, the RIXS intensity $I(\omega)$ is $I(\omega)=\sum_{f}\left| A_{f} \right|^{2}\delta({\omega}-E_{f}+E_{0})$ with $\omega$ being the incident X-ray energy, where the scattering amplitude is 
\begin{equation}
A_{f} = \sum_{m}\frac{\bra{f}\hat{D}_{\bm{\epsilon}}^{\dagger}\ket{m}H_{\rm I}\bra{m}\hat{D}_{\bm{\epsilon}}\ket{0}}{{\omega} + H_{0} -H_{\rm I} + i\Gamma},
\end{equation}
with $\hat{D}_{\epsilon}={\bm{\epsilon}}{\cdot}\bm{p}^{\dagger}s$ being the dipole operator for the $1s~{\rightarrow}~2p$ transition, $\bm{p} = (p_{x}, p_{y})$, and $s$ being the electron operator for $s$-orbit. Here, $\bm{\epsilon}$, $\bm{p}^{\dagger}$, and $s$ are the polarization of the incident X-ray,  the vector operator that creates an electron in the $2p$ orbital, and the operator that annihilates an electron in the $1s$ orbital, respectively. The propagator is $\hat{G} = ({\omega} + H_{0} -H_{\rm I} + i\Gamma)^{-1}$ when the Hamiltonians of the ground state and the intermediate states are $H_0$ and $H_{\rm I}$, respectively, and $\Gamma$ is the inverse lifetime of the core hole.
Following Ueda's paper \cite{UedaNature2023}, the intermediate Hamiltonians without involving spin interactions is  
\begin{equation} 
H^{\alpha}_{\rm I} = {\alpha}\sum_{{\langle}i,j{\rangle}}ss^{\dagger}(p_{\phi_{+}}^{\dagger}p_{\phi_{+}}-p_{\phi_{-}}^{\dagger}p_{\phi_{-}}),
\end{equation}
where
\begin{equation}
\begin{aligned}
p^{\dagger}_{\phi_{-}} &= -p^{\dagger}_{x}\sin{\phi}+p^{\dagger}_{y}\cos{\phi}\\ 
p^{\dagger}_{\phi_{+}} &= p^{\dagger}_{x}\cos{\phi}+p^{\dagger}_{y}\sin{\phi}.
\end{aligned}
\end{equation}

\subsection{The intermediate Hamiltonian involving spin interactions}

O $K$-RIXS pumps electrons from $s$-orbit to $p$-orbit. Before pumping,  
the system is also governed by the Hamiltonian with the Dzyaloshinskii-Moriya (DM) interaction for setting oxgen atoms into some displacement from helical spins:
\begin{equation} \label{Eq1}
-\lambda\sum_{{\langle}i,j{\rangle}}\left(\delta \bm{u}_{0} {\cross}\bm{e}_{\rm Ni-Ni}\right)\cdot\left(\bm{S}_i {\cross}\bm{S}_{j}\right),
\end{equation}
where $i$ and $j$ are two Ni sites connected by an oxgen atom in between 
and $\delta \bm{u}_{0}$ is the additional (but static) displacement due to non-vanishing chiral spins. $\bm{e}_{\rm Ni-Ni}$ is the unit vector along Ni-Ni. Approximate Ni-O-Ni on the same circle, then $\delta \bm{u}_{0}$ is about along the radial direction of a circle formed by the helical structure of Ni-O-Ni. $\delta \bm{u}_{0} {\cross}\bm{e}_{\rm Ni-Ni}$ is about the same direction of
$\bm{S}_i {\cross}\bm{S}_{j}$  
with the magnitude being $\delta \bm{u}_{0}({\phi})$. Here ${\phi}$
is the polar angle that describes the circle. $\delta \bm{u}_{0}({\phi})$  can be decomposed into Fourier series and in the lowest order nontrivial term, it is a constant as  $\delta \bm{u}_{0}({\phi})=\delta \bm{u}_{0}$. (Refs. \cite{OkamotoPRL2007,KatsuraPRL2007,Sergienko2006}). 
Note that the DM interaction is mediated by the oxgen atom in between.

To describe the intermediate Hamiltonian after RIXS pumping, we first remark
that the mechanism for the generation of phonon is that the electron from $s$-orbit may go into oxgen atom in either $p_{\phi_{+}}$ orbit or $p_{\phi_{-}}$ orbit, in which $\phi$ and $\delta \bm{u}$ become dynamical and thus generate the corresponding
phonon.  If the electron is pumped into $p_{\phi_{+}}$  orbit, it is governed by the interaction: 
\begin{equation} \label{Eq2}
-\lambda_{1}\sum_{{\langle}i,j{\rangle}}\left(\delta \bm{u} {\cross}\bm{e}_{\rm Ni-Ni}\right)\cdot\left(\bm{S}_i {\cross}\bm{S}_{j}\right),
\end{equation}
where $\delta \bm{u}$ is the dynamical displacement generated due to the occupation of $p_{\phi_{+}}$ orbit in the intermediate state, i.e., $\delta \bm{u}$ is the phonon  displacement superimposed on $\delta \bm{u}_0$.

Similarly, if the electron is pumped into $p_{\phi_{-}}$ orbit, it is governed by the interaction: 
\begin{equation} \label{Eq3}
-\lambda_{2}\sum_{{\langle}i,j{\rangle}}\left(\delta \bm{u} {\cross}\bm{e}_{\rm Ni-Ni}\right)\cdot\left(\bm{S}_i {\cross}\bm{S}_{j}\right)
\end{equation}

In the following, we shall choose the helical (spiral) direction to be along the $z$-direction ($c$ direction) 
\begin{equation} \label{Eq4}
\bm{S}_{i} = S^{x}_{q}\cos(qz)\hat{x}+S^{y}_{q}\sin(qz)\hat{y}
\end{equation}

\begin{equation} \label{Eq5}
\begin{aligned}
\bm{S}_{i} {\cross} \bm{S}_{j=i+1} &= \left(S^{x}_{q}\cos{q(ia)}\hat{x}+S^{y}_{q}\cos{q(ia)}\hat{y} \right){\cross} \left(S^{x}_{q}\cos{q\left((i+1)a\right))}\hat{x}+S^{y}_{q}\cos{q(ia)}\hat{y} \right) \\  &=S^{x}_{q}S^{y}_{q}\hat{z}\left[\cos{q(ia)}\sin{q\left((i+1)a\right)}-\sin{q(ia)}\cos{q\left((i+1)a\right)}\right]  \\ &= S^{x}_{q}S^{y}_{q}\sin{qa\hat{z}}
\end{aligned}
\end{equation}
For helical spins, spins are rotated by ${\pi}/2$. One gets the same result. Hence
\begin{equation} \label{Eq6}
-\lambda\sum_{{\langle}i,j{\rangle}}\left(\delta \bm{u} {\cross}\bm{e}_{\rm Ni-Ni}\right)\cdot\left(\bm{S}_i {\cross}\bm{S}_{j}\right) \propto \delta uS^{x}_{q}S^{y}_{q}\sin{qa}
\end{equation}
Now generally, we have 
\begin{equation} \label{Eq7}
\bm{S}_{i}=\bm{S}_{q}e^{i\bm{q}\cdot \bm{r}_{i}}+\bm{S}_{-q}e^{-i\bm{q}\cdot \bm{r}_{i}} 
\end{equation}

\begin{equation} \label{Eq8}
\bm{S}_{i} {\cross} \bm{S}_{j} = \bm{S}_{q} {\cross} \bm{S}_{-q} \left[e^{i\bm{q}\cdot (\bm{r}_{i} - \bm{r}_{J})} - e^{-i\bm{q}\cdot (\bm{r}_{i} - \bm{r}_{J})} \right] 
\end{equation}
Hence
\begin{equation} \label{Eq9}
-\lambda\sum_{{\langle}i,j{\rangle}}\left(\delta \bm{u} {\cross}\bm{e}_{\rm Ni-Ni}\right)\cdot\left(\bm{S}_i {\cross}\bm{S}_{j}\right) \propto \delta u \left(\bm{S}_{q} {\cross}\bm{S}_{-q} \right)_{z}
\end{equation}
So all together, in the leading order term (neglect additional $\delta \bm{u}$ displacement) and keep $\phi$ as dynamical, the intermediate Hamiltonian derived from the DM is

\begin{equation} \label{Eq10}
\begin{aligned}
H^{\rm DM}_{\rm I} = -\lambda_{1}\sum_{{\langle}i,j{\rangle}}ss^{\dagger}p_{\phi_{+}}^{\dagger}p_{\phi_{+}}^{ }\left(\delta \bm{u}_{0} {\cross}\bm{e}_{\rm Ni-Ni}\right)\cdot\left(\bm{S}_i {\cross}\bm{S}_{j}\right)  \\ -\lambda_{2}\sum_{{\langle}i,j{\rangle}}ss^{\dagger}p_{\phi_{-}}^{\dagger}p_{\phi_{-}}^{ }\left(\delta \bm{u}_{0} {\cross}\bm{e}_{\rm Ni-Ni}\right)\cdot\left(\bm{S}_i {\cross}\bm{S}_{j}\right),
\end{aligned}
\end{equation}
where $p_{\phi_{+}}^{\dagger}p_{\phi_{+}}$ describes occupation of  $p_{\phi_{+}} $ orbit and 
$p_{\phi_{-}}^{\dagger}p_{\phi_{-}}$  describes occupation of $p_{\phi_{-}} $ orbit. When $p_{\phi_{+}}$ orbit is occupied by 
the extra electron, $\lambda_{1}$ starts to work; similarly, only when $p_{\phi_{-}}$ orbit  is occupied by the extra electron, $\lambda_{2}$ starts to work. Note that here we follow Ueda’s paper \cite{UedaNature2023}
and use $\delta \bm{u}_0$ for $ H_{\rm I}$, which is the intermediate Hamiltonian before the phonon is excited.

The above intermediate Hamiltonian indicates the generating chiral phonon process: The intermediate state will generate chiral phonons dictated by the helicity of spins on Ni atoms. The parameter $\lambda$ is usually large in multiferroic materials (NTO) and hence the effect is large and exhibited as enhancement. Furthermore, for collinear spins $\bm{S}_{i}{\cross}\bm{S}_{j}=0$, $H^{\rm DM}_{\rm I}$ vanishes. 

\subsection{RIXS scattering amplitude}

Following Ueda’s paper \cite{UedaNature2023}, the RIXS scattering amplitude is 
\begin{equation} \label{Eq11}
A_{f} \propto \bra{f} D^{\dagger}_{\epsilon^{\prime}}H_{\rm I}D_{\epsilon} \ket{0},
\end{equation}
where $H_{\rm I} = H^{\alpha}_{\rm I} +H^{\rm DM}_{\rm I} $.

\vspace{3mm}

\noindent To derive $H^{\rm DM}_{\rm I}$, we first find
\begin{equation} \label{Eq13}
\begin{aligned}
\bm{\epsilon}'^{*}\cdot pp^{\dagger}_{x}p^{ }_{x}p^{\dagger}\cdot \bm{\epsilon}\ket{0} &= \epsilon'^{*}_{x}\epsilon_{x}\ket{0}  \\ \bm{\epsilon}'^{*}\cdot pp^{\dagger}_{y}p^{ }_{y}p^{\dagger}\cdot \bm{\epsilon}\ket{0} &= \epsilon'^{*}_{y}\epsilon_{y}\ket{0}  \\ \bm{\epsilon}'^{*}\cdot pp^{\dagger}_{x}p^{ }_{y}p^{\dagger}\cdot \bm{\epsilon}\ket{0} &= \epsilon'^{*}_{x}\epsilon_{y}\ket{0}  \\ \bm{\epsilon}'^{*}\cdot pp^{\dagger}_{y}p^{ }_{x}p^{\dagger}\cdot \bm{\epsilon}\ket{0} &= \epsilon'^{*}_{y}\epsilon_{x}\ket{0}.
\end{aligned}
\end{equation}
Therefore,
\begin{equation} \label{Eq14}
\begin{aligned}
&\bm{\epsilon}'^{*}\cdot pp^{\dagger}_{\phi_{-}}p^{ }_{\phi_{-}}p^{\dagger}\cdot \bm{\epsilon} \\ &= \left[\sin^{2}{\phi}\epsilon'^{*}_{x}\epsilon^{ }_{x}+\cos^{2}{\phi}\epsilon'^{*}_{y}\epsilon^{ }_{y}-\sin{\phi}\cos{\phi}\left(\epsilon'^{*}_{x}\epsilon^{ }_{y}+\epsilon'^{*}_{y}\epsilon^{ }_{x} \right)\right]\ket{0} \\ &= \left(-\epsilon'^{*}_{x}\sin{\phi}+\epsilon'^{*}_{y}\cos{\phi} \right)\left(-\epsilon^{ }_{x}\sin{\phi}+\epsilon^{ }_{y}\cos{\phi} \right)\ket{0} \\ &\equiv \epsilon'^{*}_{\phi_{-}}\epsilon^{ }_{\phi_{-}}\ket{0}.
\end{aligned}
\end{equation}
Here $\epsilon_{\phi_{-}}=-\epsilon_{x}\sin{\phi}+\epsilon_{y}\cos{\phi}=\bm{\epsilon}\cdot\left(-\sin{\phi}, \cos{\phi}\right)$.

\noindent{Similarly,}
\begin{equation}\label{Eq15}
\begin{aligned}
&\bm{\epsilon}'^{*}\cdot pp^{\dagger}_{\phi_{+}}p^{ }_{\phi_{+}}p^{\dagger}\cdot \bm{\epsilon} \\ &= \left(\epsilon'^{*}_{x}\cos{\phi}+\epsilon'^{*}_{y}\sin{\phi} \right)\left(\epsilon^{ }_{x}\cos{\phi}+\epsilon^{ }_{y}\sin{\phi} \right)\ket{0} \\ 
&\equiv \epsilon'^{*}_{\phi_{+}}\epsilon^{ }_{\phi_{+}}\ket{0}.
\end{aligned}
\end{equation}
Here $\epsilon_{\phi_{+}}=\epsilon_{x}\cos{\phi}+\epsilon_{y}\sin{\phi}=\bm{\epsilon}\cdot\left(\cos{\phi}, \sin{\phi}\right)$.
Hence, the corresponding scattering amplitude is
\begin{equation} \label{Eq16}
\begin{aligned}
A^{\rm DM}_{f} &\propto \bra{f} D^{\dagger}_{\epsilon^{\prime}}H_{\rm I}D_{\epsilon} \ket{0} \\ &\propto \left(\bm{S}_{q}{\cross}\bm{S}_{-q}\right)_{z}\delta u_{0}\left[\lambda_{1}\bra{m}
\epsilon^{{\prime}*}_{\phi_{+}}\epsilon_{\phi_{+}}\ket{0}+\lambda_{2}\bra{m}
\epsilon^{{\prime}*}_{\phi_{-}}\epsilon_{\phi_{-}}\ket{0}\right].
\end{aligned}
\end{equation}
Here $\bm{S}_{q}{\cross}\bm{S}_{-q}$ represents the helicity of spin, the polarizations of photon (before scattering and after scattering) are represented by $\epsilon$ and $\epsilon^\prime$, which also gives rise to $\phi$ dependence through $\epsilon_{\phi_{-}}$ and $\epsilon_{\phi_{+}}$, depending on the angular moment m being explored. 

\begin{equation} \label{Eq17}
\begin{aligned}
\epsilon_{\phi_{+}} &= \frac{1}{2}\left[\left(\epsilon_{x}+i\epsilon_{y} \right)(\cos{\phi}-i\sin{\phi})+\left(\epsilon_{x}-i\epsilon_{y} \right)(\cos{\phi}+i\sin{\phi}) \right]  \\ 
&=\frac{1}{2}\left(\epsilon_{L}e^{-i\phi}+\epsilon_{R}e^{i\phi} \right)  \\ 
\epsilon_{\phi_{-}} &= \frac{1}{2}\left[\left(\epsilon_{x}+i\epsilon_{y} \right)(-\sin{\phi}-i\cos{\phi})+\left(\epsilon_{x}-i\epsilon_{y} \right)(-\sin{\phi}+i\cos{\phi}) \right] \\ 
&=\frac{1}{2}\left(-\epsilon_{L}ie^{-i\phi}+\epsilon_{R}ie^{i\phi} \right)
\end{aligned}
\end{equation}
Hence
\begin{equation}\label{Eq18}
\begin{aligned}
\epsilon'^{*}_{\phi_{+}}\epsilon^{ }_{\phi_{+}} &= \frac{1}{4}\left(\epsilon'^{*}_{R}e^{-i\phi}+\epsilon'^{*}_{L}e^{i\phi}\right)\left(\epsilon^{ }_{L}e^{-i\phi}+\epsilon^{ }_{R}e^{i\phi}\right)
\end{aligned}
\end{equation}

\noindent{Similarly,}
\begin{equation}\label{Eq19}
\begin{aligned}
\epsilon'^{*}_{\phi_{-}}\epsilon^{ }_{\phi_{-}} &= \frac{1}{4}\left(\epsilon'^{*}_{L}ie^{i\phi}-\epsilon'^{*}_{R}ie^{-i\phi}\right)\left(-\epsilon^{ }_{L}ie^{-i\phi}+\epsilon^{ }_{R}ie^{i\phi}\right) \\ 
&= \frac{1}{4}\left(\epsilon'^{*}_{L}e^{i\phi}-\epsilon'^{*}_{R}e^{-i\phi}\right)\left(\epsilon^{ }_{L}e^{-i\phi}-\epsilon^{ }_{R}e^{i\phi}\right)
\end{aligned}
\end{equation}
Hence 
\begin{equation}\label{Eq20}
\begin{aligned}
&\epsilon'^{*}_{\phi_{+}}\epsilon^{ }_{\phi_{+}} - \epsilon'^{*}_{\phi_{-}}\epsilon^{ }_{\phi_{-}}\\ 
&= \frac{1}{4}\left(\epsilon'^{*}_{R}e^{-i\phi}+\epsilon'^{*}_{L}e^{i\phi}\right)\left(\epsilon^{ }_{L}e^{-i\phi}+\epsilon^{ }_{R}e^{i\phi}\right) - \frac{1}{4}\left(\epsilon'^{*}_{L}e^{i\phi}-\epsilon'^{*}_{R}e^{-i\phi}\right)\left(\epsilon^{ }_{L}e^{-i\phi}-\epsilon^{ }_{R}e^{i\phi}\right) \\ &=\frac{1}{2}\epsilon'^{*}_{R}\epsilon^{ }_{L}e^{-i2\phi}+\frac{1}{2}\epsilon'^{*}_{L}\epsilon^{ }_{R}e^{i2\phi} 
\end{aligned}
\end{equation}
This gives rise to the result in Ueda's paper \cite{UedaNature2023},
\begin{equation}\label{Eq21}
\epsilon'^{*}_{R}\epsilon^{ }_{L}e^{-i2\phi}+\epsilon'^{*}_{L}\epsilon^{ }_{R}e^{i2\phi}=
\begin{pmatrix}
\epsilon'^{*}_{R} &\epsilon'^{*}_{L}
\end{pmatrix}
\begin{pmatrix}
     0 & e^{-i2\phi}   \\
     e^{i2\phi}  & 0 
\end{pmatrix}
\begin{pmatrix}
     \epsilon_{R}  \\
     \epsilon_{L}
\end{pmatrix}
=\bm{\epsilon}'^{*}_{c}
\begin{pmatrix}
     0 & e^{-i2\phi}   \\
     e^{i2\phi}  & 0 
\end{pmatrix}
\bm{\epsilon}^{ }_{c}
\end{equation}

\noindent For our case, we first note that
\begin{equation}\label{Eq22}
\begin{aligned}
&\bra{m}\epsilon'^{*}_{\phi_{-}}\epsilon^{ }_{\phi_{-}}\ket{0}=\frac{1}{4}\bra{m}\left(\epsilon'^{*}_{L}e^{i\phi}-\epsilon'^{*}_{R}e^{-i\phi}\right)\left(\epsilon^{ }_{L}e^{-i\phi}-\epsilon^{ }_{R}e^{i\phi}\right)\ket{0} \\ 
&= \frac{1}{4}\left[\epsilon'^{*}_{R}\epsilon^{ }_{R}\bra{m}\ket{0}-\epsilon'^{*}_{R}\epsilon^{ }_{L}\bra{m}e^{-i2\phi}\ket{0}-\epsilon'^{*}_{L}\epsilon^{ }_{R}\bra{m}e^{i2\phi}\ket{0}+\epsilon'^{*}_{L}\epsilon^{ }_{L}\bra{m}\ket{0} \right]  \\
&\bra{m}\epsilon'^{*}_{\phi_{+}}\epsilon^{ }_{\phi_{+}}\ket{0}=\frac{1}{4}\bra{m}\left(\epsilon'^{*}_{R}e^{-i\phi}+\epsilon'^{*}_{L}e^{i\phi}\right)\left(\epsilon^{ }_{L}e^{-i\phi}+\epsilon^{ }_{R}e^{i\phi}\right)\ket{0} \\ 
&= \frac{1}{4}\left[\epsilon'^{*}_{R}\epsilon^{ }_{R}\bra{m}\ket{0}+\epsilon'^{*}_{R}\epsilon^{ }_{L}\bra{m}e^{-i2\phi}\ket{0}+\epsilon'^{*}_{L}\epsilon^{ }_{R}\bra{m}e^{i2\phi}\ket{0}+\epsilon'^{*}_{L}\epsilon^{ }_{L}\bra{m}\ket{0} \right]
\end{aligned}
\end{equation}
So,
\begin{equation}\label{Eq23}
\begin{aligned}
A^{\rm DM}_{f} &\propto \bra{f} D^{\dagger}_{\epsilon^{\prime}}H_{\rm I}D_{\epsilon} \ket{0} \\ 
&\propto \left(\bm{S}_{q}{\cross}\bm{S}_{-q}\right)_{z}\delta u_{0}\left[-\lambda_{1}\bra{f}\epsilon'^{*}_{R}\epsilon^{ }_{L}e^{-i2\phi}+\epsilon'^{*}_{L}\epsilon^{ }_{R}e^{i2\phi}\ket{0}+\lambda_{2}\bra{f}\epsilon'^{*}_{R}\epsilon^{ }_{L}e^{-i2\phi}+\epsilon'^{*}_{L}\epsilon^{ }_{R}e^{i2\phi}\ket{0}\right]  \\ 
&= (\lambda_{2}-\lambda_{1})\left(\bm{S}_{q}{\cross}\bm{S}_{-q}\right)_{z}\delta u_{0}\bra{f}\bm{\epsilon}'^{*}_{c}
\begin{pmatrix}
     0 & e^{-i2\phi}   \\
     e^{i2\phi}  & 0 
\end{pmatrix}
\bm{\epsilon}_{c}\ket{0}, 
\end{aligned}
\end{equation}
where in addition to the dependence on the angular momentum operator $L_{\pm}=e^{{\pm}i\phi}$, it is also proportional to the spin helicity order along the spiral/helicity direction (set as z-axis). 

Therefore, the total scattering amplitude is
\begin{equation}
\begin{aligned}
A_{f} &= A^{\alpha}_{f}+A^{\rm DM}_{f}\\
&=\left[\alpha+\left(\bm{S}_{q}{\cross}\bm{S}_{-q}\right)_{z}(\lambda_{2}-\lambda_{1})\delta u_{0}\right]\bra{f}\bm{\epsilon}'^{*}_{c}
\begin{pmatrix}
     0 & e^{-i2\phi}   \\
     e^{i2\phi}  & 0 
\end{pmatrix}
\bm{\epsilon}_{c}\ket{0}
\end{aligned}
\end{equation}

Note that the above result is due to the leading term of ${\delta}u_{0}(\phi)$ in which ${\delta}u_{0}(\phi)$ is approximated by ${\delta}u_{0}$ without $\phi$ dependence.  In this case, only angular momentum $(L_{\pm})^2$ are involved. When higher order Fourier components are included, higher order angular momentum such as $\left(L_{\pm}\right)^{3}$ will also get involved. In that case, one expects that the RIXS scattering amplitude can be written as
\begin{equation}\label{Eq24}
\begin{aligned}
A^{c}_{f} \propto &\bra{f} D^{\dagger}_{\epsilon^{\prime}}H_{\rm I}D_{\epsilon} \ket{0}\\  
\propto&\left[\alpha+\left(\bm{S}_{q}{\cross}\bm{S}_{-q}\right)_{z}(\lambda_{2}-\lambda_{1})\right]\\
&\bra{f}\bm{\epsilon}'^{*}_{c}\left[a_1
\begin{pmatrix}
     0 & e^{-i\phi}   \\
     e^{i\phi}  & 0 
\end{pmatrix}
+a_{2}
\begin{pmatrix}
     0 & e^{-i2\phi}   \\
     e^{i2\phi}  & 0 
\end{pmatrix}
+a_{3}
\begin{pmatrix}
     0 & e^{-i3\phi}   \\
     e^{i3\phi}  & 0 
\end{pmatrix}
+\cdots
\right]\bm{\epsilon}^{ }_{c}\ket{0} 
\end{aligned}
\end{equation}

\pagebreak

\section{DFT calculations}
\subsection{Phonon animation}

\begin{figure}[h!]
\centering 
\includegraphics*[width=1\columnwidth]{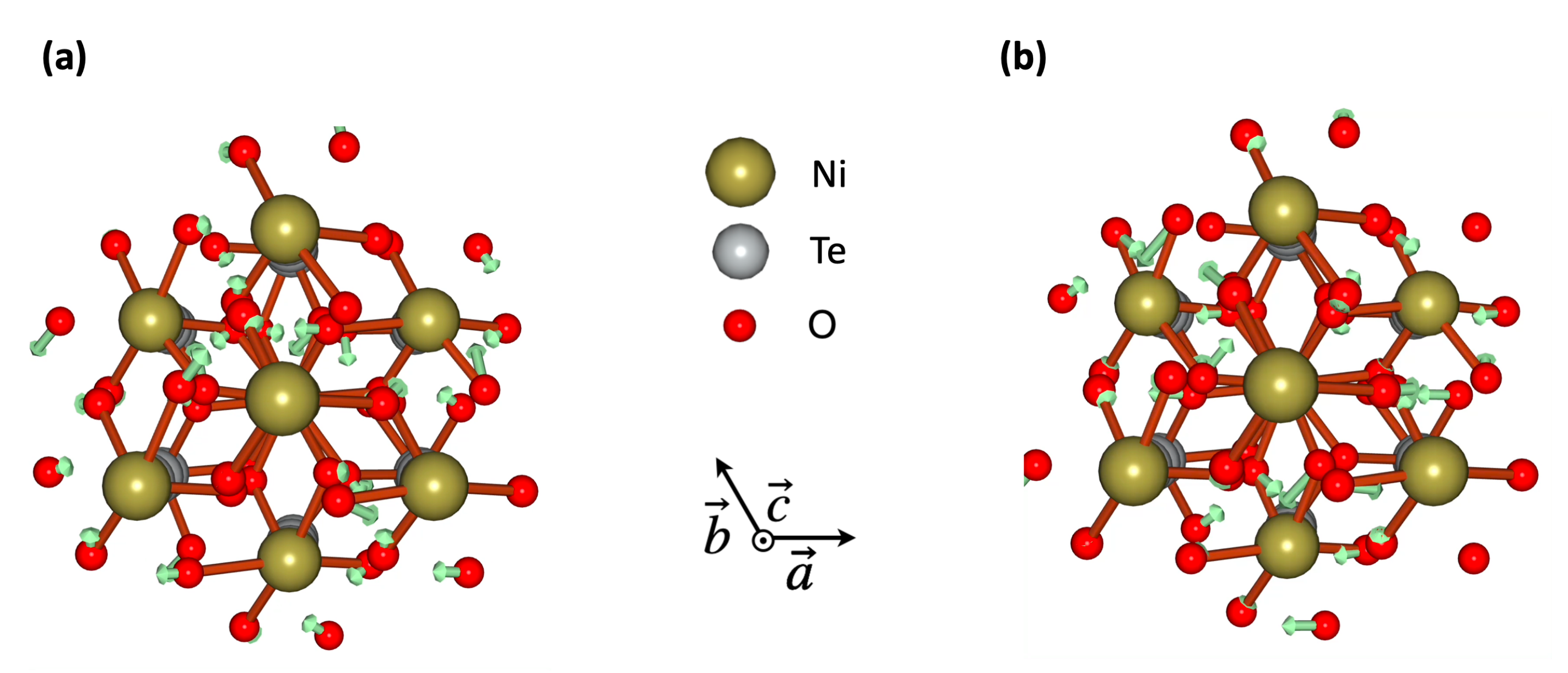}
\caption[short fig name]{{\bf Still images of phonon animations.} Panels (a) and (b) show still images of the phonon animations for the highest- and second-highest–energy phonon modes plotted in Fig. 5, corresponding to negative and positive circular phonon polarizations, respectively. These animations illustrate the collective rotation of O ions in the chiral phonon modes obtained from the GGA+U calculations for the doped system with a collinear spin structure. The arrows denote atomic displacements for each considered mode, given by the eigenvectors of the dynamical matrix. 
The animations in (a) and (b) correspond to Supplementary Video 1 and Supplementary Video 2. Here are additional links to \href{https://drive.google.com/file/d/10B1TbNPcSOxIBxtBq-0rT6qDlmlu4Awc/view?usp=share_link}{{\color{blue}animations (a)}} and \href{https://drive.google.com/file/d/1kA5YYTxdeP_SEtbU3jVzt-1y0tP4mB7P/view?usp=sharing}{{\color{blue}animations (b)}}.}
\label{phonon_animation}
\end{figure}


\bibliographystyle{naturemag}
\bibliography{ref_MnNTO}